# Electron Beam Weibel Instability in the Collisionless Shock with Low Mach Number

Jiansheng Yao[*], Yingkui Zhao[*], Biyao Ouyang, Difa Ye,

[1]Institute of Applied Physics and Computational Mathematics, Beijing 100088, China

**Correspondence to:** Jiansheng Yao; Yingkui Zhao

**Email**: ab135794@mail.ustc.edu.cn; zhao_yingkui@iapcm.ac.cn

**Abstract**

   The electron beam Weibel instability in the electrostatic collisionless shock is studied via particle in cell simulation. When the non-relativistic incoming plasmas collide with cold dilute plasmas, an electrostatic shock forms near the interface. Following that, a filamentary out-of-plane magnetic field is formed as a result of the Weibel instability.

It is demonstrated that the anisotropy of incoming hot electrons is insufficient to trigger the Weibel instability. And the Weibel instability is excited by cold electrons in dilute plasmas. After being accelerated to relativistic velocities by the shock electric field into the dense plasmas, electrons in the dilute plasma have considerable anisotropy and can trigger the Weibel instability

## 1. Introduction:

   The Weibel instability (Fried 1959; Weibel 1959), has been thought to be of great importance to the study of magnetic field amplification and gamma-ray bursts (GRBs), active galactic nuclei, pulsar wind nebulae, and supernova remnants in astrophysical environments (Blandford and Eichler 1987; Piran 2005; Waxman 2006). According to previous study (Medvedev and Loeb 1999; Nishikawa et al., 2005; Stockem et al., 2014), this instability is generally caused by the anisotropic electron distribution. Previous simulations (Mora 2005; Mora and Grismayer 2009; Thaury et al., 2010) have proved that the electrons of plasmas expanding into vacuum lose kinetic energy along expansion direction. This makes electrons exhibit obvious anisotropy, which will trigger the Weibel instability. However, when a plasma expands into a dilute plasma (Sarri et al., 2011), the Weibel instability will be suppressed due to the superthermal electrons from dilute plasmas reducing the thermal anisotropy. Further investigation (Bret et al., 2014; Artem Bohdan et al., 2021; Fiuza et al., 2020) illustrated that when two counterstreaming plasmas penetrate into each other with relativistic velocity, a shock is formed and strong magnetic fields triggered by ion-ion beam Weibel instability emerge at the shock transition.

Compared with relativistic counterstreaming plasma beams in their studies, non-relativistic plasma beams penetrating background plasmas are more common in astronomical observations. Through the simulation of the non-relativistic electron-proton plasma flow triggered by the ultraintense lase pulse, previous studies (Fiuza et al., 2012; Park et al., 2015) have demonstrated that the magnetic fields associated with Weibel instability caused by streaming electrons impede the imcoming plasmas, resulting in a significant compression and the production of a shock. That is, in their condition, the formation of shock wave is after the Weibel instability triggered by cold electron beams, and it is also called Weibel-instability-mediated shock.

However, according to our simulation, the process can be reversed in our study, i.e., the formation of filamentous magnetic field corresponding to the Weibel instability would occur after the non-relativistic shock. When the non-relativistic streaming velocity of incoming plasmas exceeds the ion acoustic velocity in dilute plasmas, the ion acoustic waves in dilute plasmas stack up, and an electrostatic shock forms near the interface. The shock is the collisionless because the mean collision free path is much longer than the size of shock wave (Zhang et al., 2017). Most importantly, electrons in the dilute plasma will be accelerated by the shock electrostatic field to relativistic velocities into the dense plasma, which in turn excites the Weibel instability to generate a strong magnetic field. This indicates that, via electrostatic shock accelerating electrons to relativistic velocities, non-relativistic plasma flows can also trigger the Weibel instability. We will describe this process in detail in this paper.

The paper is organized by four sections: the simulation model is presented in the Sec. 2; the simulation results are illustrated in Sec 3; summary and discussion are presented in Sec 4.

## 2. Setup

We explore the Weibel instability in the collisionless shock using 2D kinetic simulations with relativistic particle in cell (PIC) code EPOCH (Arber et al., 2015). There are $24000 \times 400$ cells in the simulation domain. The size of simulation box is $12000\lambda_{De} \times 3.2\lambda_e$, where $\lambda_{De} \sim 1 \times 10^{-8}\ m$ and $\lambda_e \sim 3.1 \times 10^{-5}\ m$ are Debye length and inertial length of electrons in dilute plasmas. Therefore, the gird size is $\Delta x=0.5\lambda_{De}$ and $\Delta y=0.015\lambda_e$, which is sufficient to resolve the shock whose normal is along $x$-axis and magnetic fields stripes along y-aixs triggered by the Weibel instability. The two boundaries along the $x$-axis are absorbing, while the two boundaries along the y-axis are periodic. To avoid the influence of the absorbing boundary, the particles are distributed in the simulation domain with $1000\Delta x$ away from the left and right boundaries.

The initial condition is presented in the table 1, the size of simulation box is $0 < x < 120\ \mu m$, and $-100\ \mu m < y < 100\ \mu m$, dense plasmas at the left side ($10\ \mu m < x < 70\ \mu m$) flow into dilute plasmas at the right side ($70\ \mu m < x < 110\ \mu m$) with $v_d = 500\ km/s$, which is less than the ion acoustic wave velocity in dense plasmas but greater than that in dilute plasmas. Therefore, a electronstatic collisionless shock will form in the interchange face owing to ion acoustic waves in dilute plasmas stack up. A realistic mass ratio $m_i/m_e = 1836$ is applied in our simulation, the time step is $\Delta t = 0.9\Delta x/c = 1.5 \times 10^{-17}\ s$ to satify the Courant–Friedrichs–Lewy condition, and we use 100 particles per cell to reduce numerical error. According to previous study (Ross et al 2012), the ion-ion mean free path is estimated to be:

$$\lambda_{ii}(cm) \simeq 5\times10^{-13} \frac{A_Z^2}{Z^4} \frac{[U(cm\ s^{-1})]^4}{n_Z(cm^{-3})}, \tag{1}$$

Where $A_z$, Z and $n_Z$ are the averaged atomic weight, charge, and ion number density, and U is the beam velocity of the streaming ions. Substituting these values shown in table 1, we will obtain $\lambda_{ii} \sim 3 \times 10^{-5}\ m$, while the size of electrostatic shock is serval $\lambda_{De} \sim 1 \times 10^{-8}\ m$, therefore, the ion-ion mean free path is much larger than the size of electrostatic shock, thus the shock is collisionless ((Ross J.S. et al 2012; Fiuza et al., 2012; Park et al., 2015). We have tested different simulation box sizes and resolutions to ensure convergence of the results.

**Table 1**. Parameters of plasmas

|  | Dense plasmas | Dilute plasmas |
| --- | --- | --- |
| Number Density ($m^{-3}$) | $1 \times 10^{27}$ | $5 \times 10^{25}$ |
| Electron Temperature (eV) | 5000 | 100 |
| Ion Temperature (eV) | 100 | 100 |
| Electron Debye Length ($\mu m$) | 0.016 | 0.01 |
| Electron Inertial Length ($\mu m$) | 7.21 | 32.2 |
| Flow Velocity ($ms^{-1}$) | $5.0 \times 10^6$ | 0 |
| Ion-acoustic Wave Velocity ($ms^{-1}$) | $8.94 \times 10^6$ | $1.79 \times 10^6$ |

## 3. Results

We start by considering the shock formation. In Figure 1, we present the spatial distributions of electric field and particle densities at $t = 0.05\,ps$. In Figure 1 (a), incoming electrons in dense plasmas (red dashed line) pile up at $x = 70.1\,\mu m$, and protons in dilute plasmas (magenta solid line) also pile up around $x = 70.2\,\mu m$. Calculating the net charge $N^+ - N^-$, we get graph Figure 1 (c), in which the positive charge (protons) number density $N^+ = N_{H_l} + N_{H_r}$ ($N_{H_l}$ and $N_{H_r}$ are number densities of protons at left and right sides, respectively), the negative charge (electrons) number density $N^- = N_{e_l} + N_{e_r}$ ($N_{e_l}$ and $N_{e_r}$ are number densities of electrons at left and right sides, respectively). Obviously, there is net charge around the shock surface, which indicates that electrons and protons are separated at the shock region, and electrons and protons accumulate around $x = 70.1\,\mu m$ and $x = 70.4\,\mu m$, respectively. This charge separation would generate a strong electrostatic field $E_x$ at $x = 70.25\,\mu m$ as shown in Figure 1 (c), the field strength can reach $1.5 \times 10^{11}\,V/m$ and is distributed in a narrow range $70.2\,\mu m < x < 70.3\,\mu m$ (shown in Figure 1 (d)). It is vital to note that, accelerated by this positive electrostatic filed, electrons in dilute plasmas would penetrate into incoming dense plasmas (as illustrated by the black dashed line in Figure 1 (a)).

After the shock is formed, we will investigate the Weibel instability in the shock by investigating the evolution of the out-of-plane magnetic field $B_z$. The distributions of $B_z$ at $t = 0.05\,ps, 0.5\,ps, 3\,ps$ are shown in Figure 2 (a)-(c), respectively. Among them, Figure 2 (a) corresponds to the moment when the shock is formed. Obviously, the magnetic field filamentation doesn't occur when the shock is formed. This differs from prior research (Fiuza et al., 2012) on Weibel-instability-mediated collisionless shocks, in which the shock forms after the Weibel instability. As illustrated by Figure 2 (b), the filamentous magnetic field $B_z$ occurs at $t = 0.5\ ps$, and the wavelength is about $2.85 \times 10^{-6}\ m$. At late time, corresponding to the nonlinear evolution of the Weibel instability as shown in Figure 2 (c), the filament wavelength and field strength increase due to filament merging (Medvedev et al., 2005; Gedalin et al., 2010; Peterson et al., 2021). In the following, we shall investigate the mechanism of excitation of Weibel instability in the shock wave with low Mach number.

According to linear theory (Medvedev and Loeb 1999), the formation of filamentous magnetic field corresponding to the Weibel instability needs anisotropic distributed electrons, and the dispersion relation is as follows:

$$1 = \frac{c^2 k^2}{\omega^2} + \frac{\omega_p^2/\gamma}{\omega^2}[G(\beta_\perp) + \frac{1}{2}\frac{\beta_\parallel^2}{1-\beta_\perp^2}(\frac{k^2 c^2 - \omega^2}{\omega^2 - k^2 c^2 \beta_\perp^2})], \tag{2}$$

where $\beta_\parallel = u_\parallel/c, \beta_\perp = u_\perp/c$, $\gamma = 1/\sqrt{1 - (u_\parallel^2 + u_\perp^2)/c^2}$, $G(\beta_\perp) = (2\beta_\perp)^{-1}\ln[(1 + \beta_\perp)/(1 - \beta_\perp)]$, and $\parallel$ and $\perp$ here are parallel and perpendicular directions with respect to the shock's normal direction, respectively. As demonstrated by previous study, the Weibel instability occurs for the wavelength $k$ in the range:

$$0 < k^2 < (\frac{\omega_p^2}{\gamma^2 c^2})\frac{\beta_\parallel^2}{2\beta_\perp^2(1+\beta_\perp^2)} - G(\beta_\perp), \tag{3}$$

therefore, in order for the Weibel instability to develop, the following criteria must be met:

$$\beta_\parallel^2 > 2G(\beta_\perp)\beta_\perp^2(1-\beta_\perp^2). \tag{4}$$

For incoming electrons in dense plasmas, $\beta_\parallel^2 = 2.78 \times 10^{-6}$, while $2G(\beta_\perp)\beta_\perp^2(1-\beta_\perp^2) = 0.038$, thus the Eq. (4) is invalid and incoming hot electrons in dense plasmas couldn't trigger the Weibel instability. Interestingly, when we investigate the evolution of electrons in phase space, we find that cold electrons in dilute plasmas are accelerated to relativistic velocities. The distributions of cold electrons in dilute plasmas in phase space $x - v_x$ at $t = 0.05\ ps, 0.5\ ps, 3\ ps$ are shown in Figures 3 (a)-(c), respectively. These three moments respectively correspond to the three moments depicted in Figures 2(a)-(c). It is clear that cold electrons in dilute plasmas is accelerated to $v_x \approx -7.5 \times 10^7\ ms^{-1}$. This phenomenon has previously been discovered in simulation study (Sarri et al., 2011), but no further research has been performed. After that, at $t = 0.5\ ps$ (shown in Fig. 3 (b)), these electrons are accelerated further and pierce deeper in dense plasmas. At $t = 3\ ps$ (shown in Fig. 3 (c)), these electrons encounter the plasma sheath located at $15\ \mu m < x < 25\ \mu m$, and then they are reflected and accelerated in the opposite direction, and re-enter the shock layer eventually.

In this part, via linear theory, we will demonstrate these relativistic electrons accelerated by the shock could trigger Weibel instability. When $t = 0.5\ ps$, $u_\parallel \approx 8.\times 10^7\ ms^{-1}$ (as shown in Figure 3(b)), thus $\beta_\parallel^2 = 0.071$, while $2G(\beta_\perp)\beta_\perp^2(1 - \beta_\perp^2) = 2.6 \times 10^{-4}$, therefore, the Eq. (4) is valid and these cold electrons in dense plasmas would trigger the Weibel instability. The wavenumber corresponding to the fastest grow mode (Medvedev and Loeb 1999) is

$$k_{\max} = \frac{\omega_{pe}^2}{\gamma c^2(1-\beta_\perp^2)}[\frac{-\beta_\parallel^2}{2(1-\beta_\perp^2)} - G(\beta_\perp) + \frac{(1+\beta_\perp^2)\beta_\parallel}{\sqrt{2}(1-\beta_\perp^2)^{3/2}}(\frac{\beta_\parallel^2}{1-\beta_\perp^2} + \frac{1-2\beta_\perp^2-\beta_\perp^4}{\beta_\perp^2}G(\beta_\perp))^{1/2}],$$

(4)

where $\omega_{pe} = \sqrt{n_e e^2/m_e \varepsilon_0} \approx 1.1 \times 10^{14} s^{-1}$ is the plasma frequency of electrons in dilute plasmas. We obtain the wavenumber corresponding to the fastest grow mode $k_{\max} \approx 1.44 \times 10^6\ m^{-1}$, thus the corresponding wavelength is $\lambda_{\max} = 4.38 \times 10^{-6}\ m$, this value is larger than the wavelength in Fig. 2 (b) ($\lambda \sim 2.85 \times 10^{-6}\ m$), which could be attributed to the superposition of fluctuations with various wavelengths in addition to the wave with the fastest growth rates in Fig. 2(b). In conclusion, according to the linear theory, incoming electrons in dense plasmas having low Mach number couldn't trigger the Weibel instability, while cold electrons in dilute plasmas accelerated to relativistic velocities by electrostatic shock would trigger the Weibel instability.

In this section, we will further demonstrate above conclusion via simulation. Figures 4(a)-(c) are number densities of electrons in dilute plasmas distributed in the $x-y$ plane at $t = 0.05\ ps, 0.5\ ps, 3\ ps$, respectively. Figures 4(d)-(f) are current densities of electrons in dilute plasmas distributed in the $x-y$ plane at $t = 0.05\ ps, 0.5\ ps, 3\ ps$, respectively. These three moments respectively correspond to the three moments depicted in Figures 2 and 3 (a)-(c). Obviously, at $t = 0.05\ ps$, cold electrons in dilute plasmas are accelerated and move towards left side. After that, at $t = 0.05\ ps$ presented in Figure 4 (b), distinct filamentary structures appear in electrons' density and current distributions. This indicates these relativistic electrons generate out-of-plane filamentary magnetic fields shown in Figure 2 by forming filamentary current structures. At $t = 3\ ps$, filamentary currents merge and wavelength increases significantly. Besides, we have demonstrated that there are no filamentary structures in incoming electrons' density and current distributions via simulation (not show). In conclusion, our simulation also demonstrates that the filamentary magnetic field in Fig. 2 created by the Weibel instability is excited by cold electrons in dilute plasmas, which is accelerated to relativistic velocities by the shock wave. This is consistent with the conclusions obtained from linear theory.

## 4. Conclusion

In this paper, we have shown that when the non-relativistic incoming plasmas collide with cold dilute plasmas, an electrostatic shock forms near the interface. Because of the low anisotropy of the incoming hot electrons, the Weibel instability cannot arise. After that, electrons in the dilute plasma will be accelerated by the shock electrostatic field to relativistic velocities into the dense plasmas. These cold relativistic electrons have strong anisotropy and can trigger the Weibel instability. The strong out-of-plane magnetic field decelerates the incoming plasmas and mixes particles in this region. What's more, electrons would rotate around the magnetic field and emit synchrotron radiation.

This process can have important implications for the gamma-ray bursts and collisionless shock energy dissipation. What's more, a recent work (Bohdan et al., 2021) has established the connection between the Weibel instability and magnetic field amplification at high Mach shocks. Our work extends their research to some extent, even if the incoming plasmas have a low Mach number, the shock electric field can accelerate cold electrons to relativistic speeds, trigger the Weibel instability, and generate strong magnetic fields.

**Acknowledgments:**

This work is supported by the National Science Foundation of China 11822401, 12174034. Data underlying the results presented in this paper are not publicly available at this time but can be obtained from the authors upon reasonable request.

**Figure Captions:**

**Figure 1.** The spatial distribution of electric field and particle densities at $t = 0.05\ ps$. **(a)** particle densities distribution at $t = 0.05\ ps$, they are normalized with their respective initial densities. Electrons and protons in the figure are denoted by dashed and solid lines, respectively. The subscript $l$ indicates incoming particles at the left side of simulation domain, while the subscript $r$ represents dilute particles at the right side of simulation domain. **(b)** the electrostatic field $E_x$ distributed in the $x - y$ plane. **(c)** the net charge $N^+ - N^-$ distribution at $t = 0.05\ ps$, in which, the positive charge number density $N^+ = N_{H_l} + N_{H_r}$ ($N_{H_l}$ and $N_{H_r}$ are number densities of protons at left and right sides, respectively), the negative charge number

density $N^- = N_{e_l} + N_{e_r}$ ($N_{e_l}$ and $N_{e_r}$ are number densities of electrons at left and right sides, respectively). **(d)** the electrostatic field $E_x$ distributed along $x$, this graph is obtained by averaging graph b in the y direction.

**Figure 2.** Out-of-plane magnetic field $B_z$ distributed in the $x-y$ plane at (a) $t = 0.05\ ps$, (b) $0.5\ ps$, (c) $3\ ps$.

**Figure 3.** Cold electrons in dilute plasmas distributed in phase space $x - v_x$ at (a) $t = 0.05\ ps$, (b) $0.5\ ps$, (c) $3\ ps$.

**Figure 4.** (a)-(c) are number densities of electrons in dilute plasmas distributed in the $x - y$ plane at $t = 0.05\ ps, 0.5\ ps, 3\ ps$, respectively. (d)-(f) are current densities of electrons in dilute plasmas distributed in the $x - y$ plane at $t = 0.05\ ps, 0.5\ ps, 3\ ps$, respectively.

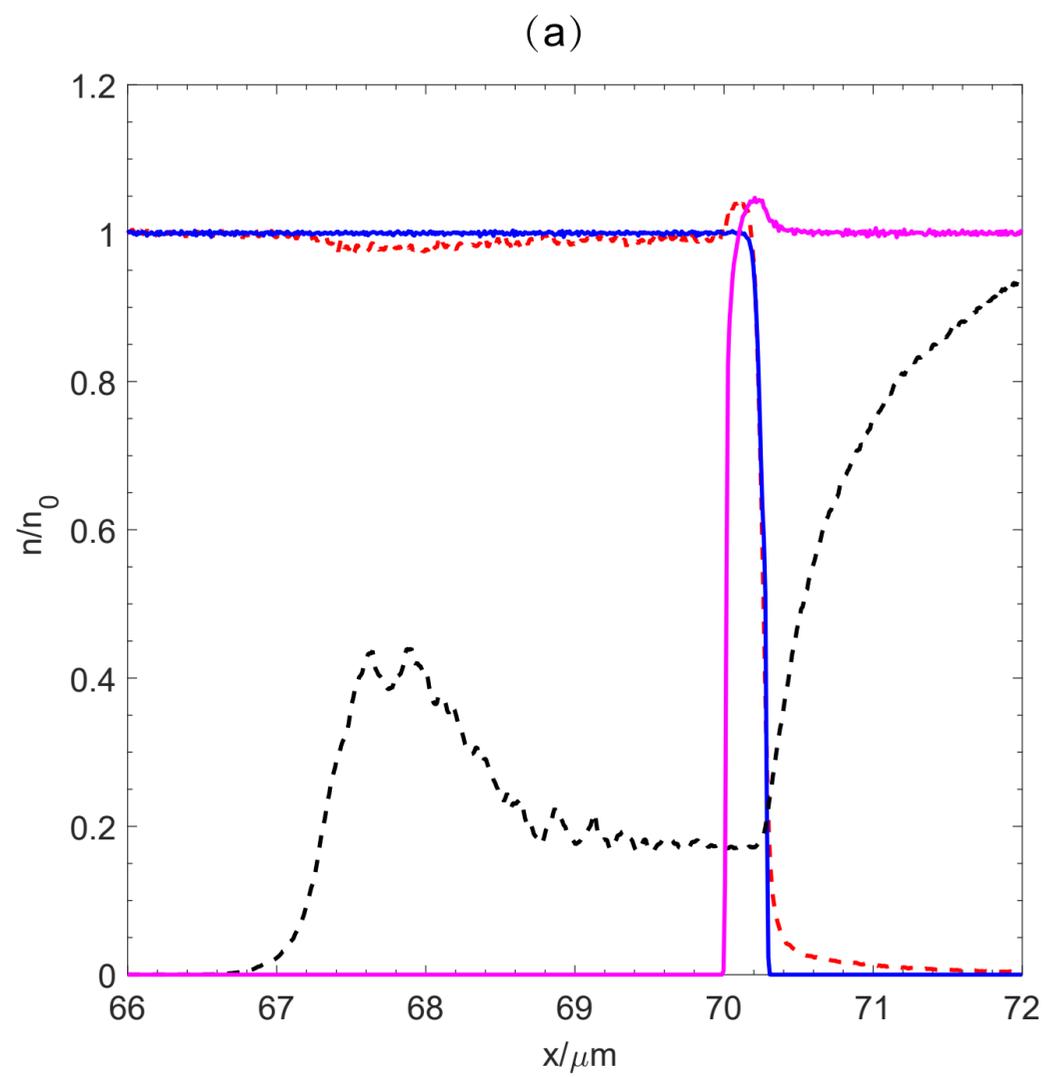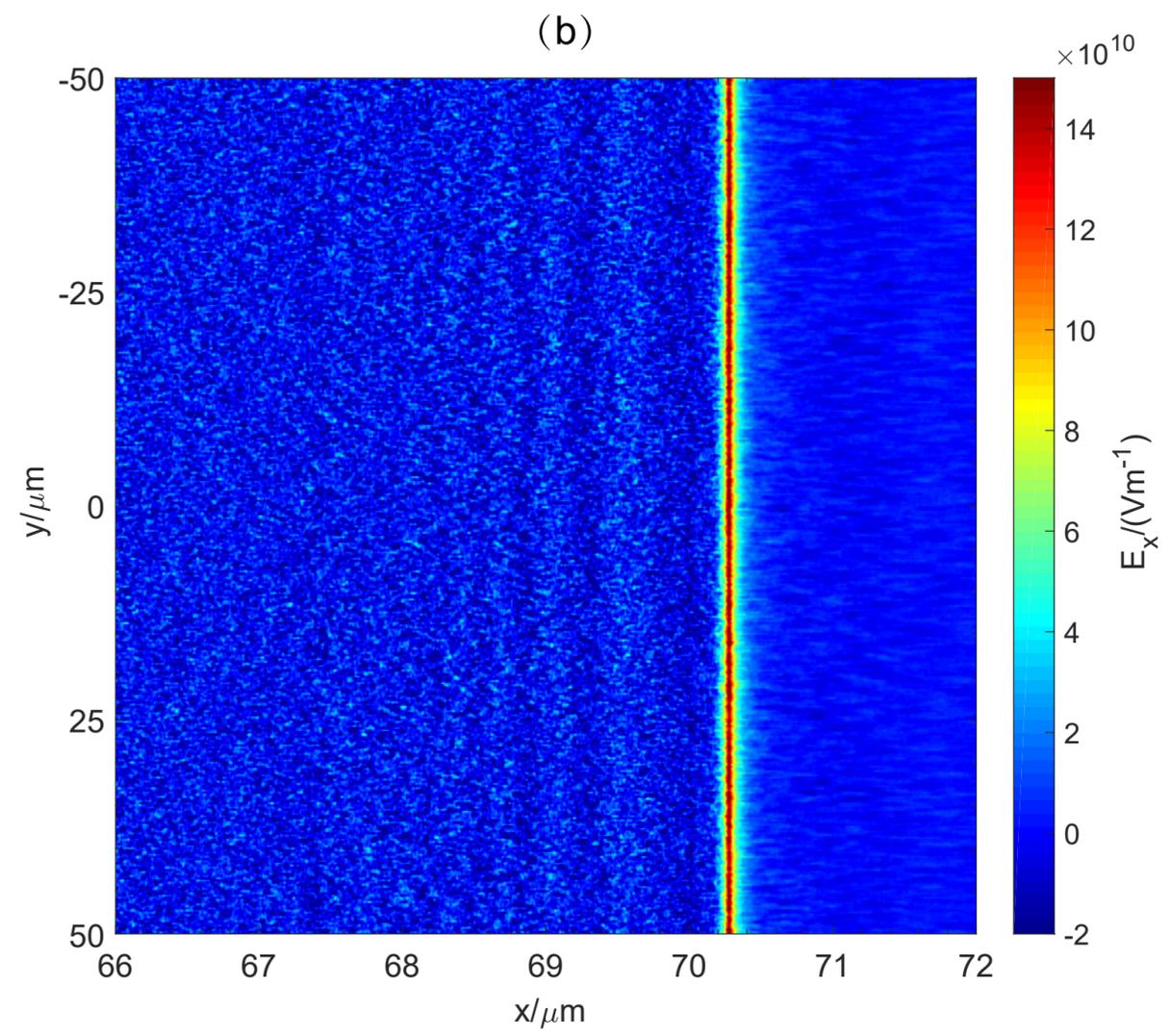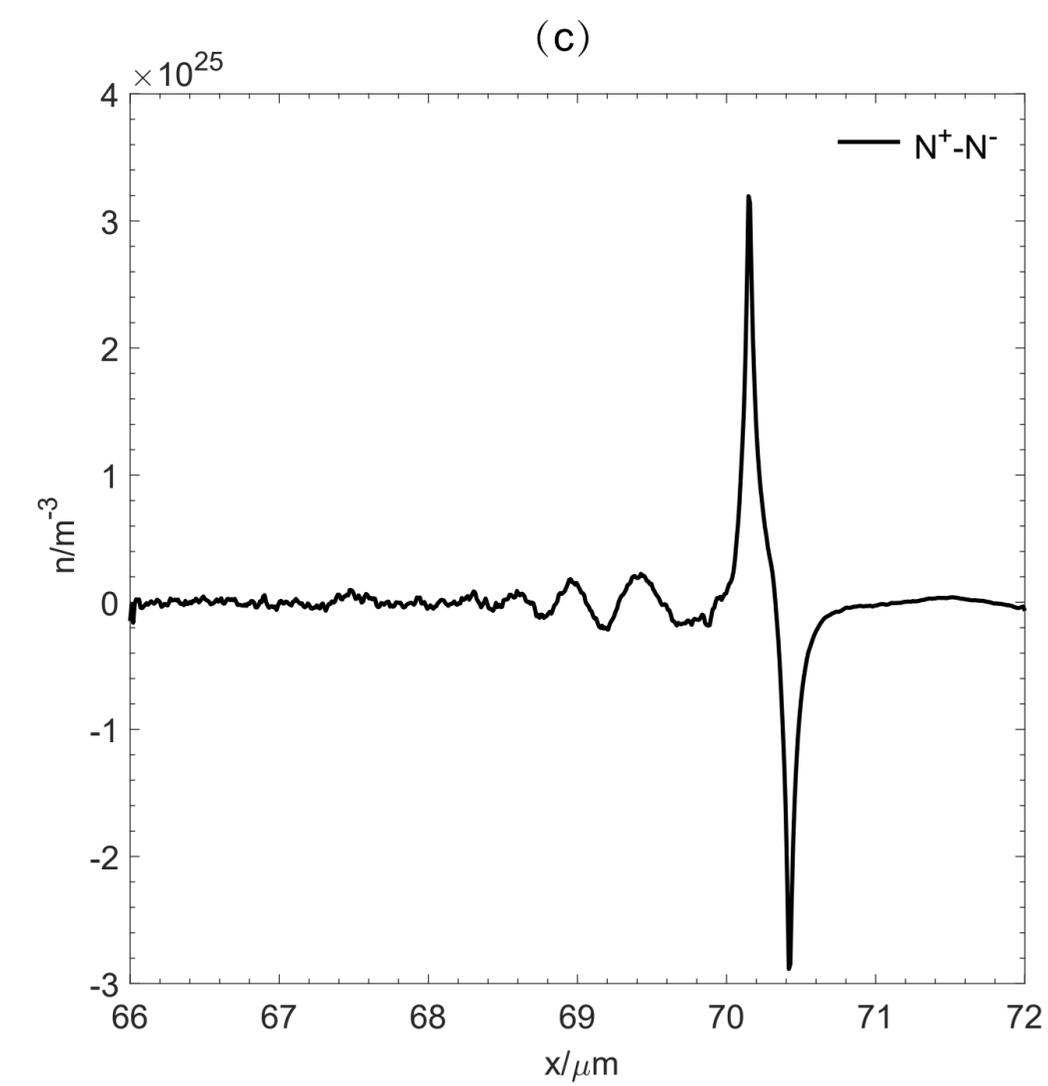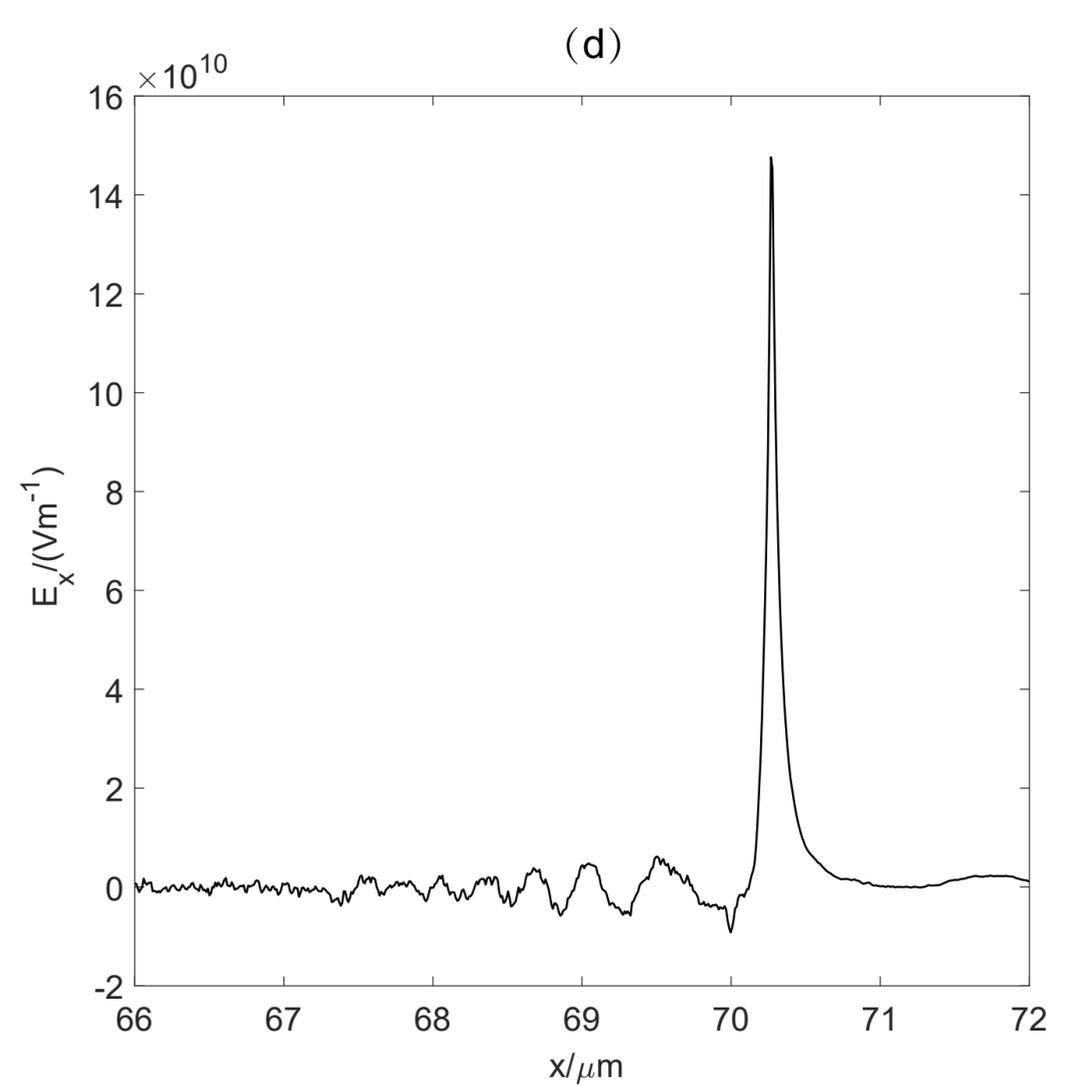

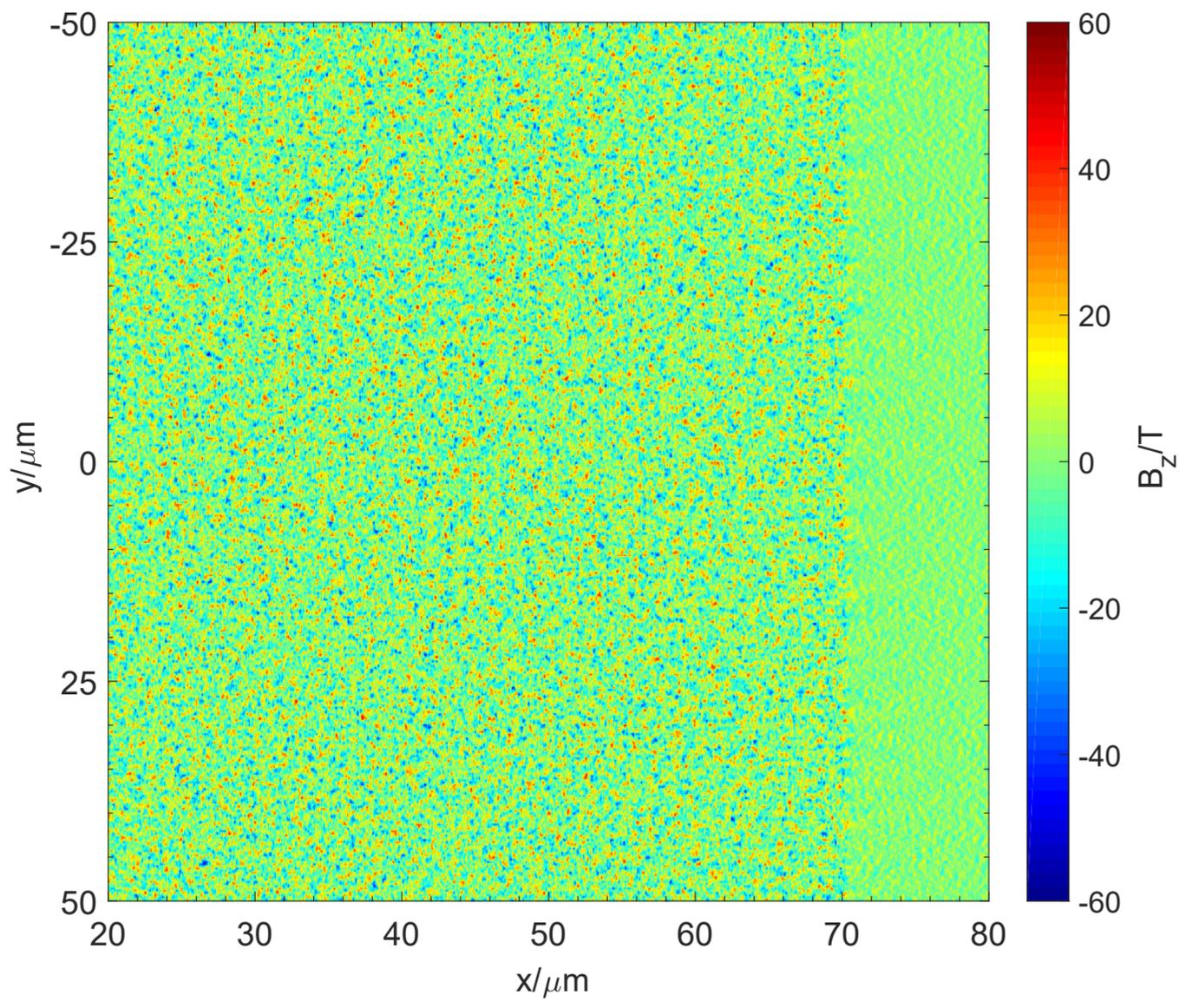 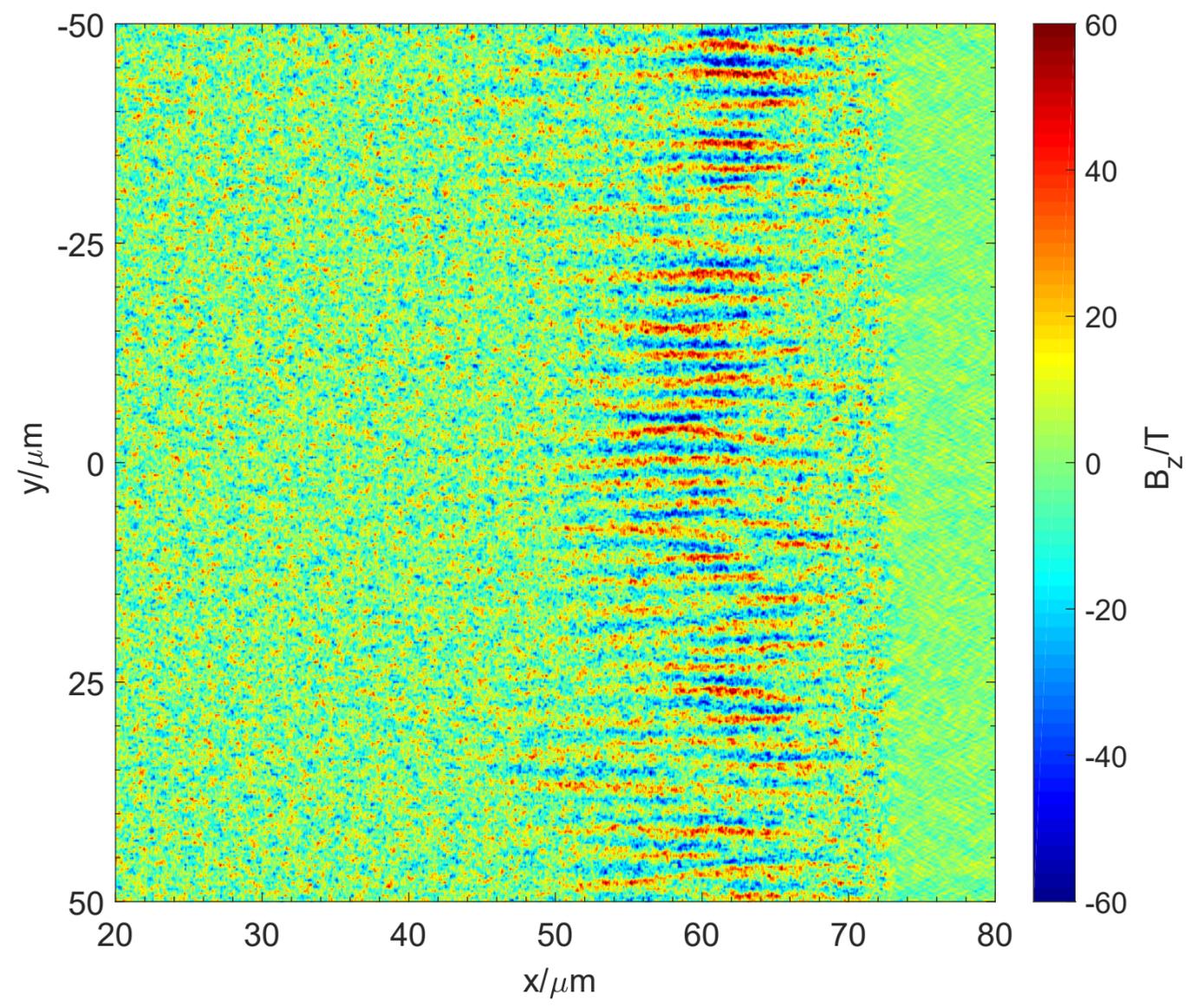 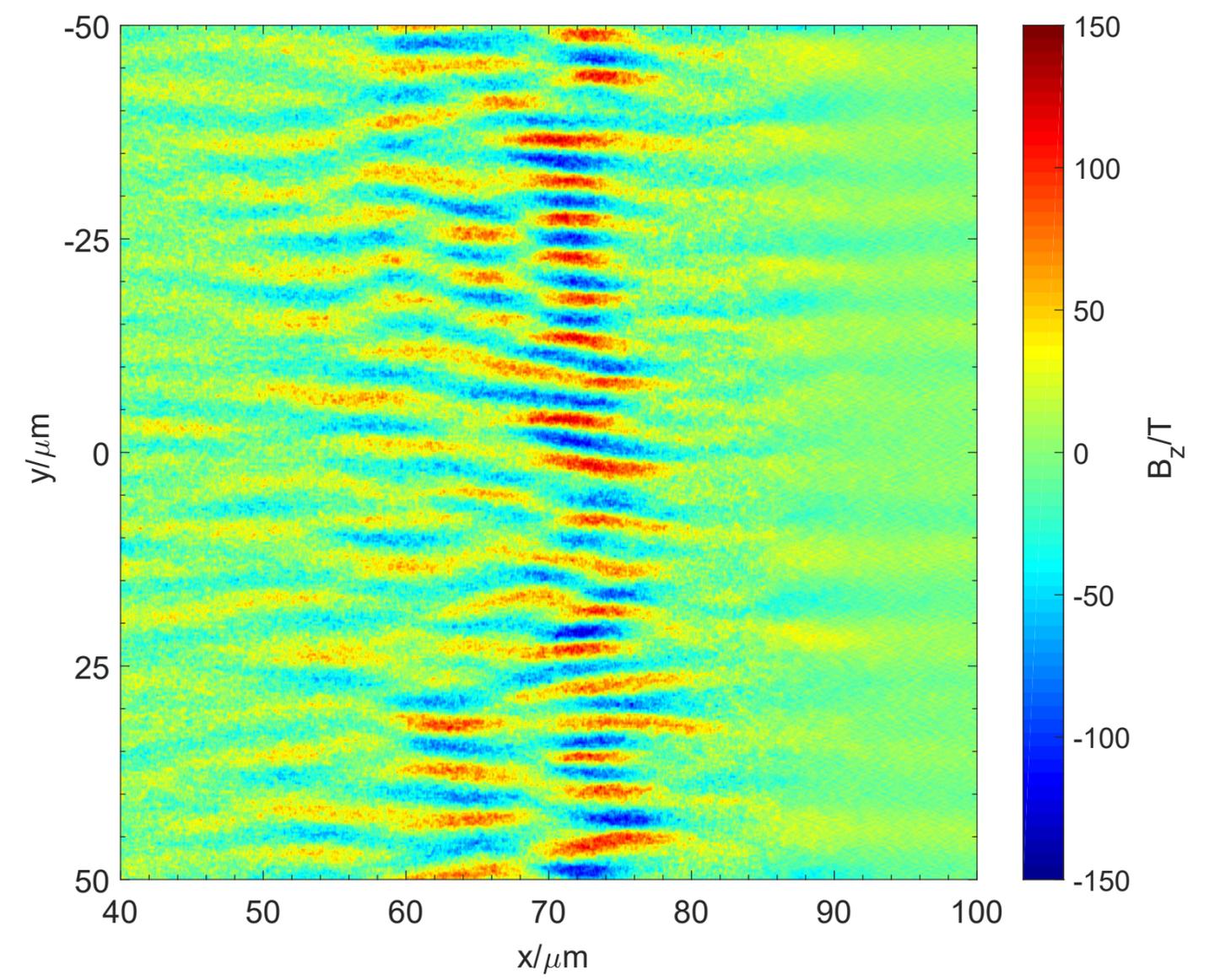

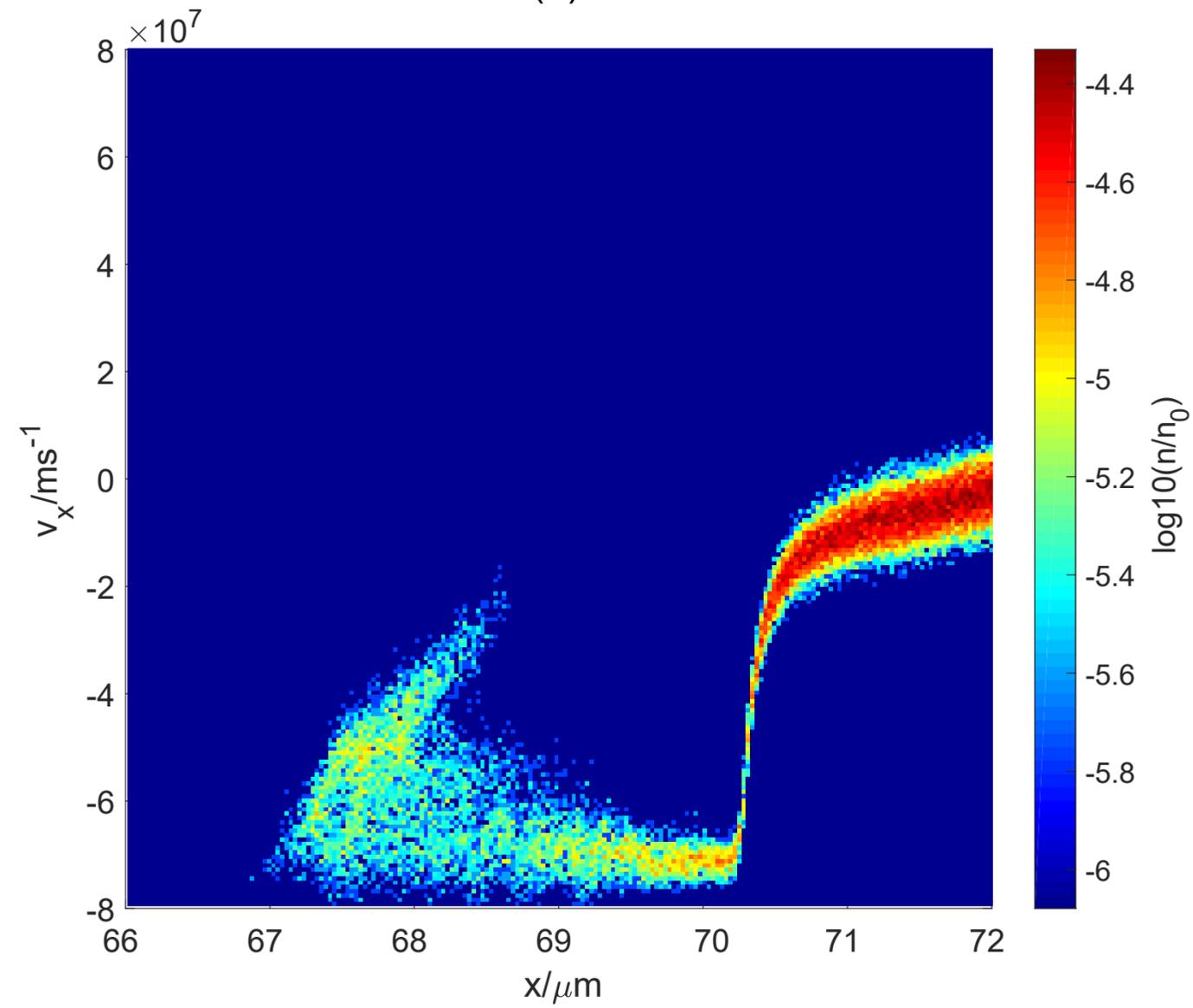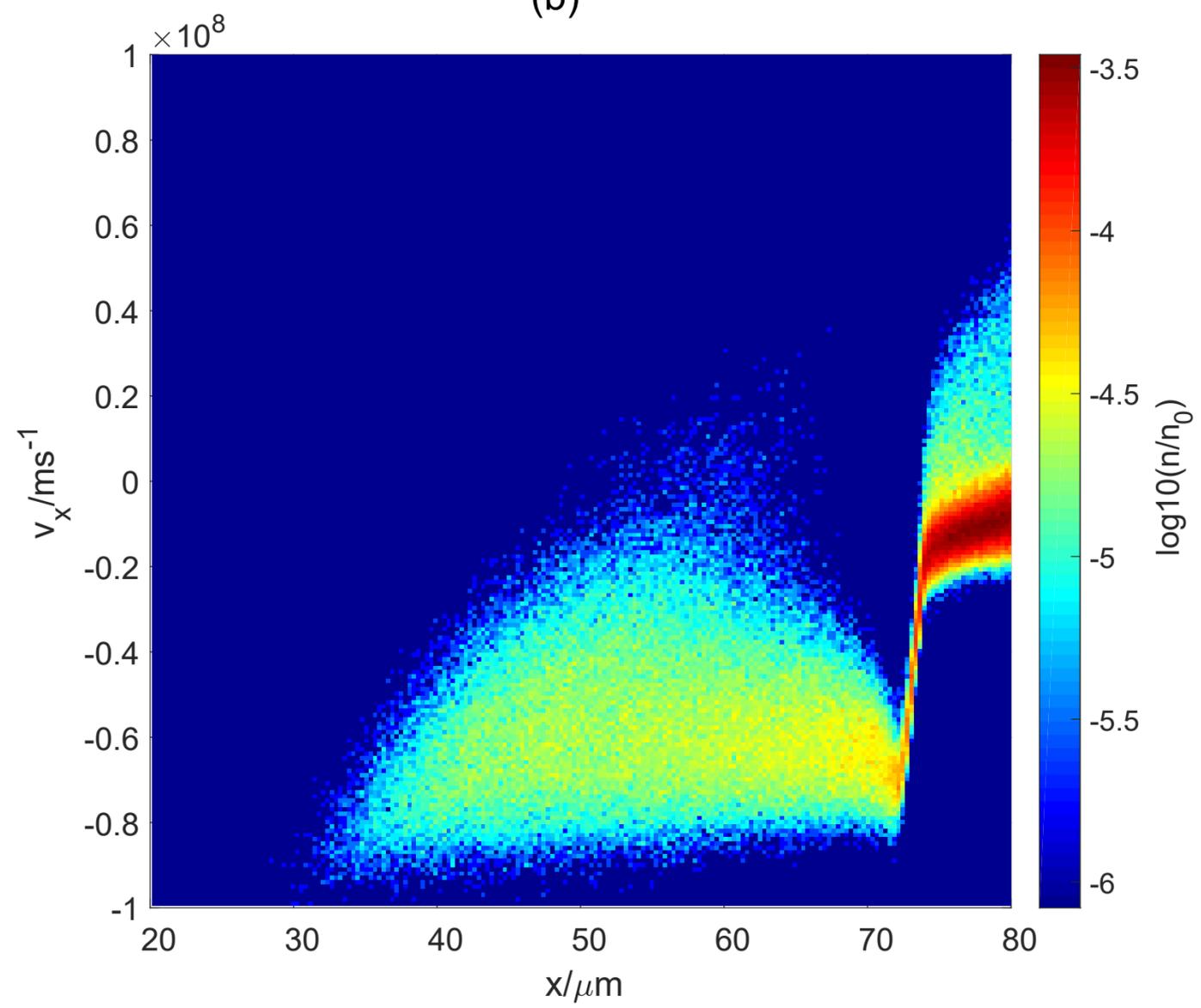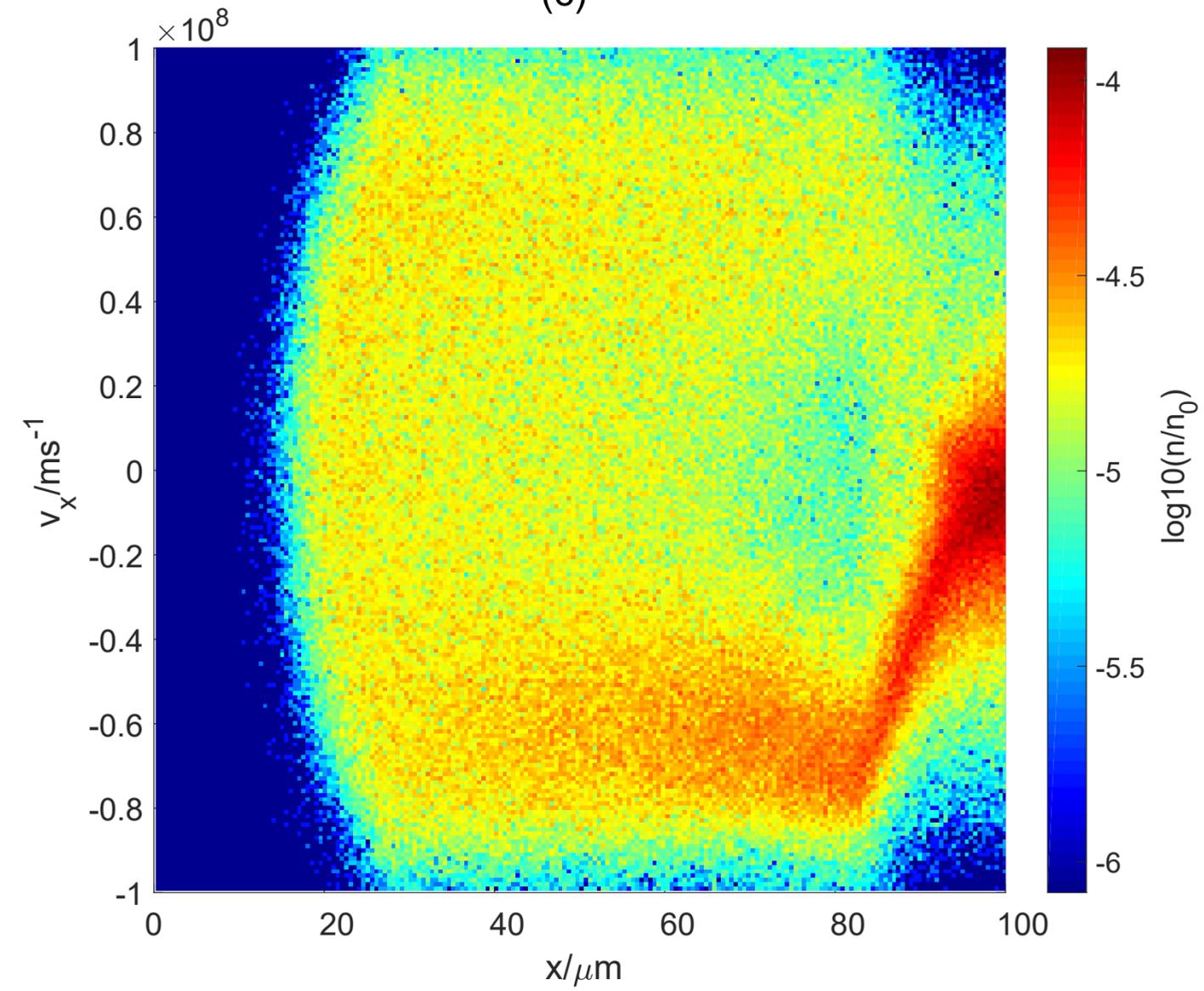

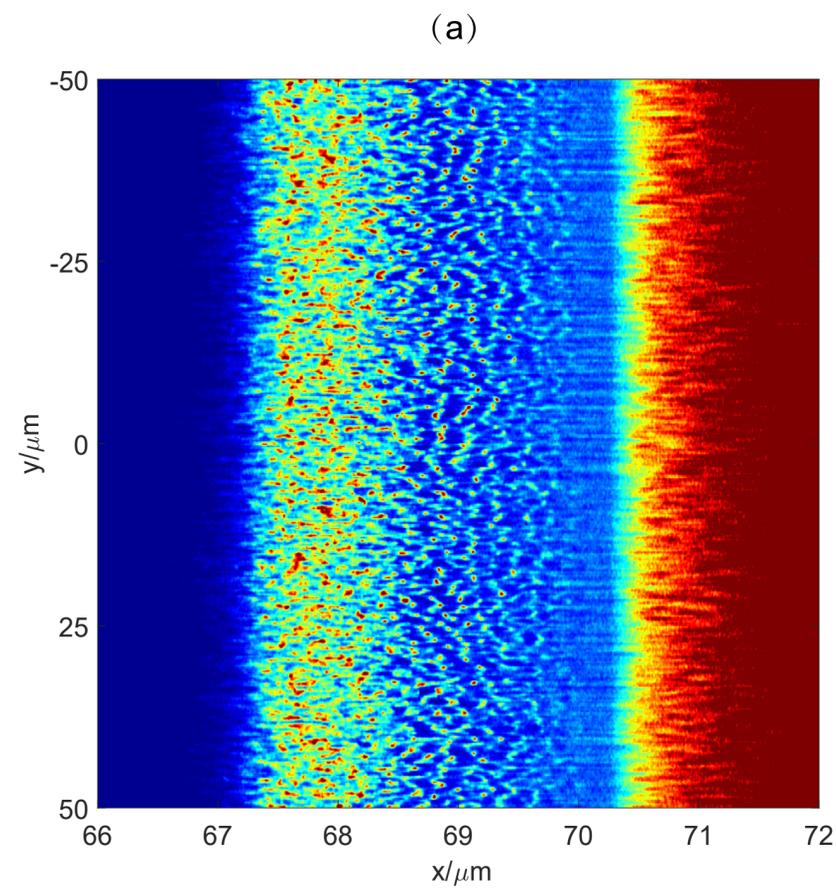 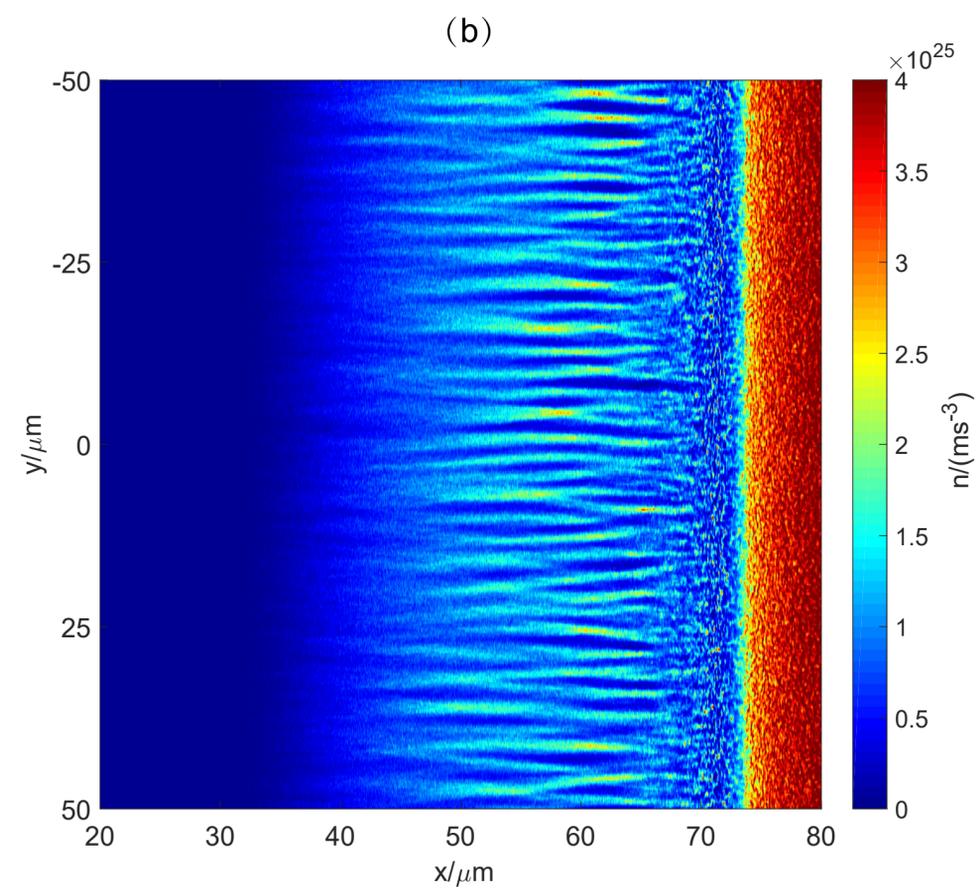 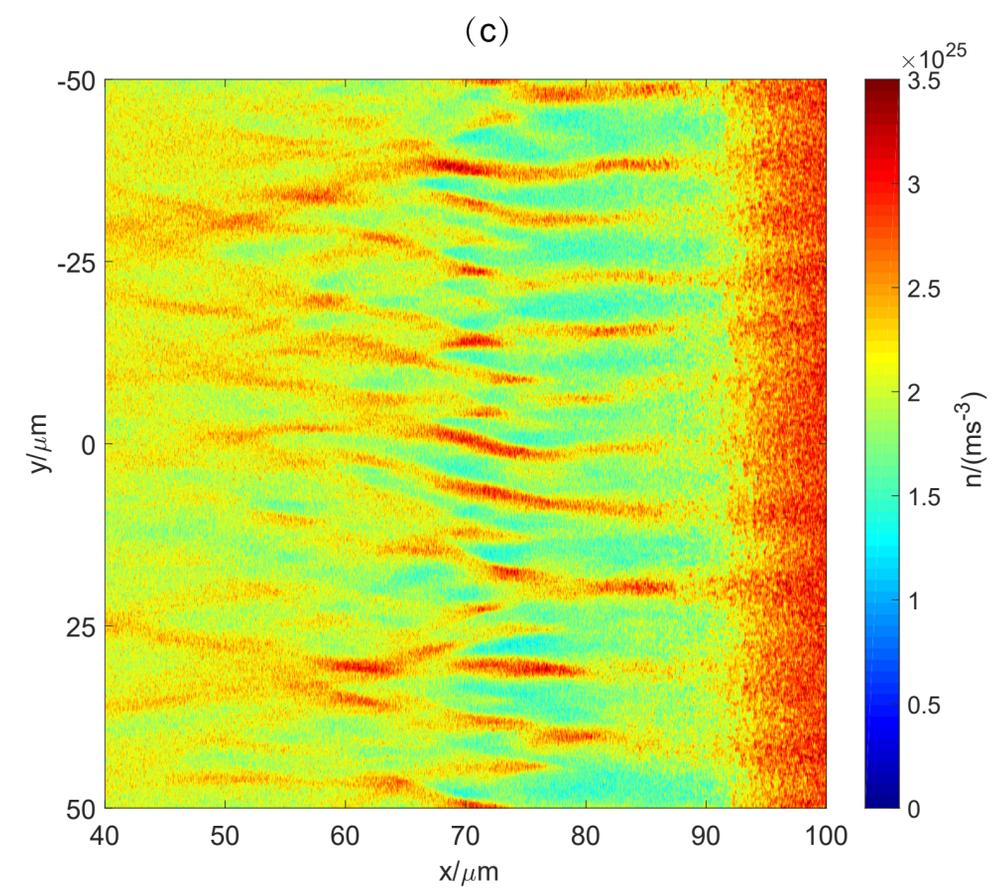
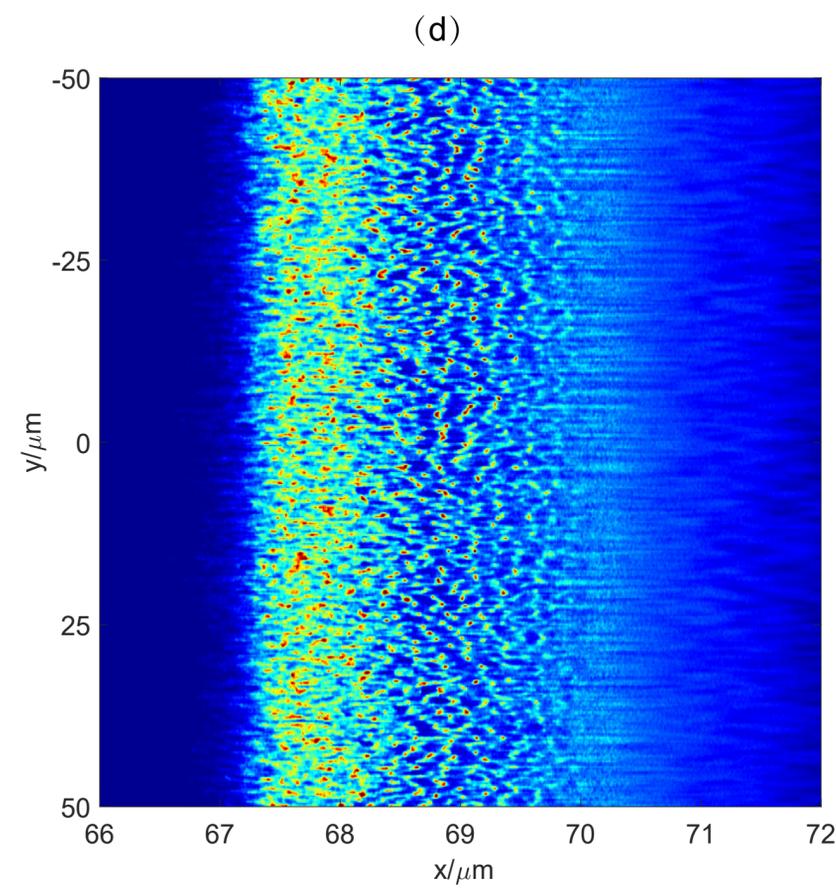 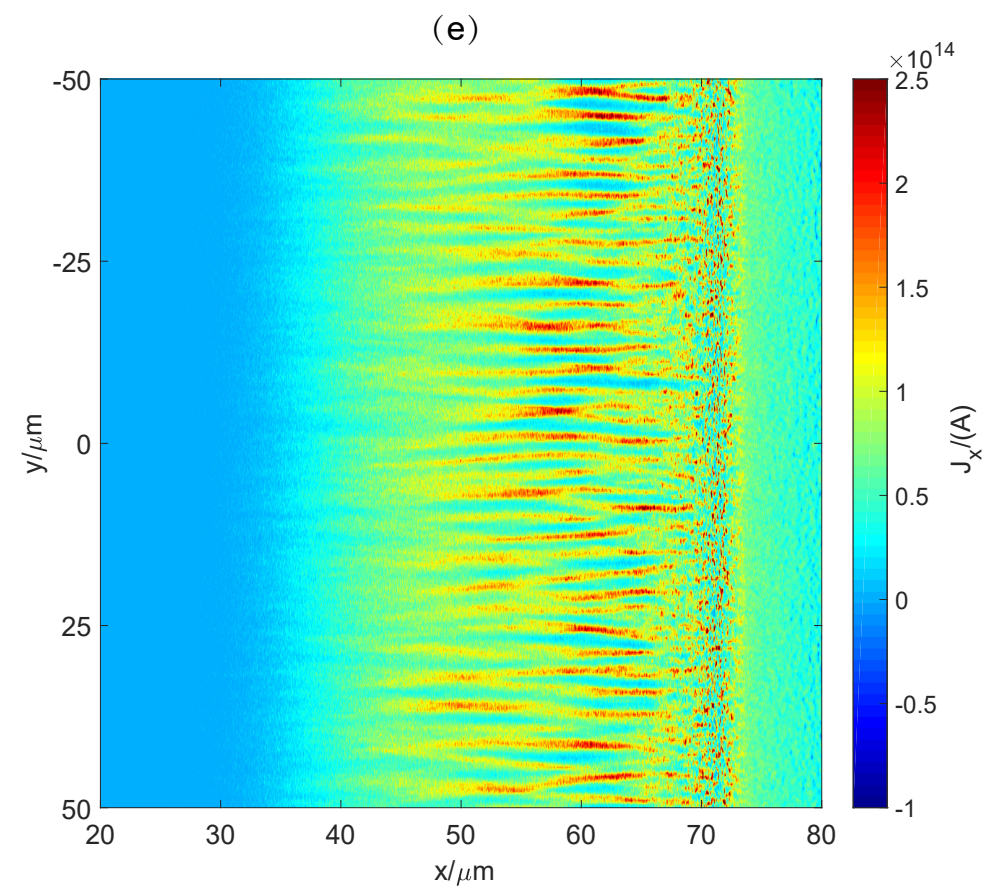 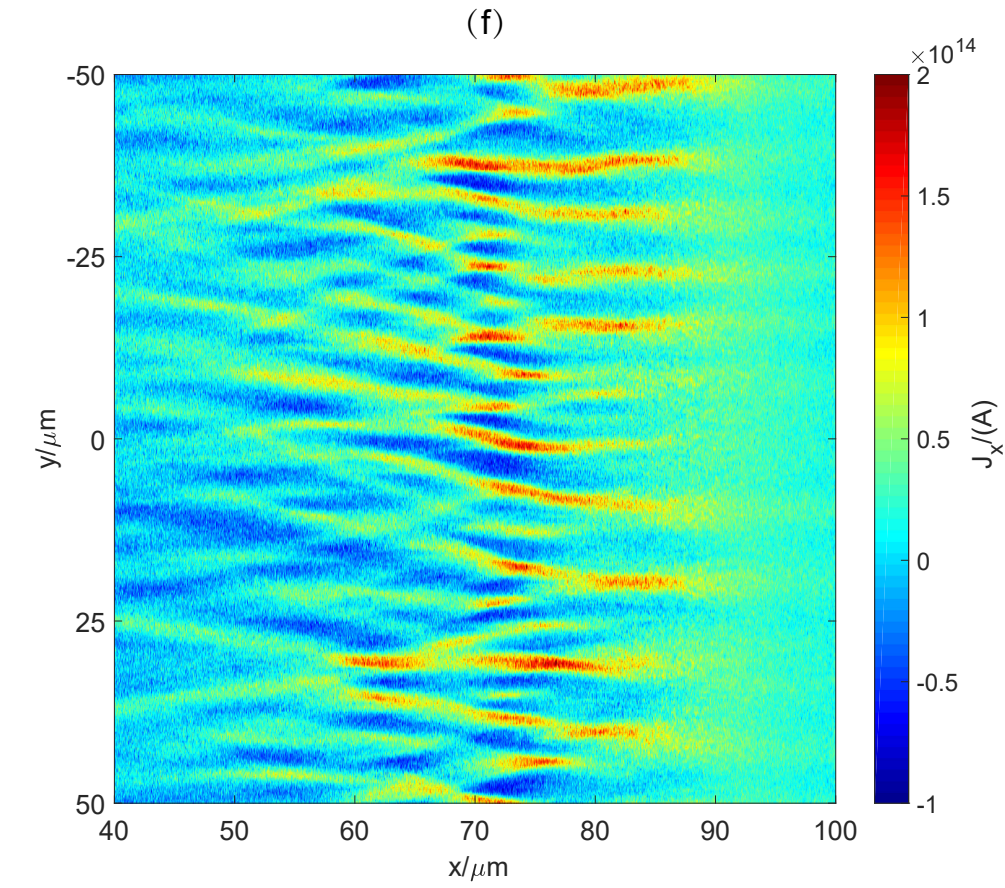